\documentclass{article}
\usepackage{amsmath}

\begin{document}

\begin{center}
Axiomatization of Mechanics

T.F. Kamalov

Physics Department, Moscow State Open University

E-mail: timkamalov@gmail.com

\end{center}

\textit{The problem of axiomatization of physics formulated by Hilbert as
early as 1900 and known as the Sixth Problem of Hilbert is nowadays even
more topical than at the moment of its formulation. Axiomatic inconsistency
of classic, quantum, and geometrized relativistic physics of the general
relativistic theory does not in the least fade away, but on the contrary,
becomes more pronounced each year. This naturally evokes the following
questions: 1. Is it possible, without drastically changing the mathematics
apparatus, to set up the axiomatics of physics so as to transform physics,
being presently a multitude of unmatched theories with inconsistent
axiomatics, into an integrated science? 2. Is it possible, maybe through
expanding their scopes, to generalize of transform the existing axiomatics
into an integral system of axioms in such a manner that existing axiomatics
of inconsistent theories would follow there from as a particular case?}



An axiom is a statement adopted without proof. Therefore, there is no other
way of obtaining an axiom than to merely guess it. It is rather hard to
resist the temptation of declaring an axiom every phenomenon inexplicable in
the framework of a particular theory. In such a case, on the one hand, the
consistency of a theory would not be compromised, and on the other hand, the
problem of explanation of this phenomenon is avoided. However, this approach
is erroneous and fallacious, as the number of axioms cannot grow in an
uncontrollable manner; the number of axioms shall be minimized. Only this
way one can expect minimization of likely errors should some axioms be
guessed inadequately. At the same time, it is impossible to avoid axioms at
all, because, according to Godel theorem, each theory comprises statements
impossible to be proved within the framework of this theory. It is these
statements that constitute the foundation of the theory, governing its
results and implications.

On the one hand, it seems evident that having altered a single word in an
axiom, we could obtain dramatic changes in the theory. Hence, arbitrary
altering of axioms is inadmissible, as otherwise we would get a chaotic set
of axioms rather than an axiomatic system. On the other hand, science is not
a church doctrine, but rather, a system of theories based on guessed axioms.
Therefore, one should be extremely careful in altering axioms to adapt them
to the obtained new results. Otherwise, the description of the physical
reality would result as a multitude of inconsistent, often mutually
contradicting, axioms. And, as it has been mentioned, the number of axioms
should be reduced to the minimum.

Historically, the axiomatics of physics has by all means experienced
alterations. For example, if we compare the classical Newtonian physics with
its predecessor, Aristotelian physics, we can easily see that while
axiomatics of Aristotelian physics presumed motion to occur only provided a
force being applied to the body, the axiomatics of Newtonian classical
physics states that motion may occur also if no force is applied to the
body. Therefore, Aristotelian physics postulates that the dynamics of bodies
is described by first order differential equations, whereas Newtonian
physics (with the Laws of Newton being from the mathematical viewpoint the
axiomatics of classical physics[1]) postulates proportionality of the force
applied to the body to its acceleration. In other words, the Laws of Newton
postulate the description of body dynamics by the second order differential
equations. Newton employed the concept of the absolute space being the space
related to fixed stars. Such space can be called Euclidian.

Is it correct to consider the Laws of Newton to be the axiomatics of
classical physics? The answer is definitely positive. Shall we, and may we,
expand their scope of application to microobjects? Quantum mechanics is a
physical description of particles employing the definition of inertial
reference frame, hence employing the First Newton's Law. The First Law
states that any body free form interactions with other bodies would have
constant velocity. So, how is velocity defined in quantum mechanics? It is
done through the average with wave function $\psi $

\begin{equation}
<\dot{x}>=\int \psi ^{\ast }x\psi dx.  \label{1}
\end{equation}

In a quantum (real) reference frame $<\dot{x}>=const$ there always exist
infinitesimal fields, waves, and forces perturbing an ideal inertial
reference frame. This follows from one of the general definitions of Mach
principle [2]: \textquotedblleft Local physical laws are determined by
large-scale structure of the Universe\textquotedblright . Commenting the
Mach principle, let us note that in this case the definition of inertial
properties of a body is determined by multi-particle interactions with all
bodies in the Universe. The description of the case of particle motion with
higher derivatives of coordinates in time has been for the first time
published in 1850 by M. Ostrogradsky; it is known as an Ostrogradsky
Canonical Formalism [3]. Being a mathematician, Ostrogradsky considered
coordinate systems rather than reference systems. This case corresponds to a
quantum (real) reference frame comprising not only inertial reference
frames, but also non-inertial ones, determined, according to Mach principle,
by multi-particle interactions with all bodies moving in the Universe.
Inertial reference frames defines by all bodies moving in the Universe [2]

\begin{center}
$\frac{d^{2}}{dt^{2}}(\frac{\int \rho (r)rdV}{\int \rho (r)dV})=0$.
\end{center}

Here $\rho (r)$ is the function of the distribution of all masses in the
Universe with the volume $V$.

\textbf{Definition}

\textit{A kinematic state of a mechanical system with constant higher
derivative }$\dot{x}^{(n)}=const$\textit{\ is called defined if the
kinematics of the body is described with a differential equation}

\begin{equation}
F(x,\dot{x},\ddot{x},\dddot{x},...,\dot{x}^{(n)})=0  \label{2}
\end{equation}%
\qquad \qquad \qquad \qquad \qquad

Let us assume that if in an arbitrary (any) reference frame the average
value of a higher derivative is constant,

\begin{equation}
<\dot{x}^{(n)}>=<\frac{d^{n}x}{dt^{n}}>=const  \label{3}
\end{equation}

then the function $F$ is finite.

\textbf{Kinematic Principle (Inertial Principle)}

\textit{A kinematic state of a mechanical system free from interactions with
other bodies is observer-dependent and persists until its interaction with
other bodies alters its kinematic state.}

The acceleration for a body with free from interactions with other bodies is
a constant for the observer in the constant-accelerated reference frame. In
this case the acceleration is define the kinematic state of the body because
in that case the acceleration is the invariant for the reference frame of
the observer.

Let us call an invariant of a reference frame the constant higher derivative
that does not change in case of transformation of coordinates

\begin{eqnarray}
x^{\prime } &=&f(x,\dot{x},\ddot{x},\dddot{x},...,\dot{x}^{(n)})  \label{4a}
\\
t^{\prime } &=&t  \label{4b}
\end{eqnarray}

\qquad \qquad \qquad \qquad \qquad

Then the kinematic state of a mechanical system free form interactions with
other bodies depends on the invariant of the observer's reference frame.

Let us call a harmonic reference system the reference system with a clock
and an observer oscillating harmonically, in which any body free from
interactions with other bodies would maintain the average value of its
higher derivative. For a reference system oscillating harmonically let us
consider the invariant the $<\dot{x}^{(n)}>=const$. In a harmonic reference
frame a coordinate of the body may be described by the function

\begin{equation}
\varphi (t,x)=\varphi _{0}\exp i(kx+\omega t)  \label{5}
\end{equation}

being $\varphi _{0}$ the amplitude of oscillation, $k$ and $\omega $ the
wave vector and angular frequency of oscillations, respectively.

In the particular case,

\begin{equation}
\varphi (t,x)\approx kx+\omega t=K_{i}X^{i}  \label{6}
\end{equation}

and being $K_{i}$ 4-dimensional wave vector and $X^{i}$ 4-coordinate of the
body, $i=0,1,2,3$\ in a harmonic reference frame. In this case, function $%
\varphi $ describes the 4-coordinate of the body multiplied by a constant
coefficient. In a vibrating reference frame the coordinate of a body may be
expressed with arrays of harmonic oscillations.

\textbf{Dynamic Principle}

\textit{There exist reference frames with a clock and an observer, in which
the dynamics of a body is described be the equation:}

\begin{equation}
k_{1}\dot{x}+k_{2}\ddot{x}+k_{3}\dddot{x},+...k_{2n}\dot{x}^{(2n)})=F(x,\dot{%
x},\ddot{x},\dddot{x},...,\dot{x}^{(n)})  \label{7}
\end{equation}

We will call such reference frames real (or quantum reference frames).

Let us call a force $F$ the quantitative measure of the interaction between
the bodies.

The generalized principle of relativity of Galileo means in this case that
the order of the differential equation (6) describing the dynamics of a
mechanical system with the invariant of the reference frame $\dot{x}%
^{(n)}=const$ does not alter at transformation (3).

The differential equation (6) corresponds to the description of the body
dynamics in a non-isolated (open) mechanical system with the external forces
of the system with odd derivatives, corresponding, for example, to losses
due to friction and radiation. Odd derivatives correspond to losses
(friction or radiation) and describe irreversible cases for open systems not
satisfying variational principles of mechanics. The case of an isolated
(close) mechanical system corresponds the differential equation with even
derivatives.

\begin{equation}
k_{2}\ddot{x}+k_{4}\ddddot{x}+...k_{2n}\dot{x}^{(2n)})=F(x,\dot{x},\ddot{x},%
\dddot{x},...,\dot{x}^{(n)})  \label{8}
\end{equation}

The reference frames, in which the dynamics of a system is described by the
equation

\begin{equation}
<k_{2}\ddot{x}>=<F(x,\dot{x},\ddot{x},\dddot{x},...,\dot{x}^{(n)})>,
\label{9}
\end{equation}

we will call inertial reference frames. Here, the proportionality
coefficient in the equation $k_{2}$ of dynamics (6) is the mass of a body.

\textbf{Static Principle}

\textit{If a particle rests along an arbitrary direction, then the resultant
force acting thereon along this direction is zero.}

For inertial frames, the Lagrangian $L$ depends only on coordinates and
their first derivatives $L=L(x,\dot{x})$ [4]. For the case of quantum (real)
reference frames the Lagrangian depends on coordinates and their higher
derivatives and has the form $L=L(x,\dot{x},\ddot{x},\dddot{x},...,\dot{x}%
^{(n)})$.

Let us consider in more detail such precise description of dynamics of
bodies motion accounting for quantum (real) reference frames determined,
according to our model, by complex multi-particle interactions with all
bodies in the Universe.

For an accurate description of dynamics of bodies motion accounting for
higher derivatives, let us consider the body in an arbitrary reference
frame, denoting the position $r$ of the body in the space as and time as $t$%
. Then, expanding the function $r=r(t)$ into Taylor's series in the zero
point, we get

\begin{equation}
r=r_{0}+\dot{r}t+\frac{1}{2!}\ddot{r}t^{2}+\frac{1}{3!}\dddot{r}t^{3}+...+%
\frac{1}{n!}\dot{r}^{(n)}t^{n}+...  \label{10}
\end{equation}

Let us denote

\begin{center}
$r_{N}=r_{0}+\dot{r}t+\frac{1}{2!}\ddot{r}t^{2}$
\end{center}

and the additional correction variables $q_{r}$ for our model with arbitrary
reference frames as , where - is a value equal to zero in the classical
Newtonian mechanics

\begin{equation}
q_{r}=\frac{1}{3!}\dddot{r}t^{3}+...+\frac{1}{n!}\dot{r}^{(n)}t^{n}+...
\label{11}
\end{equation}

Then

\begin{center}
$r=r_{N}+q_{r}.$
\end{center}

In our case the description discrepancy and \textit{uncertainty between the
two models} $h$ is equal to the difference in descriptions of a test
particle in the extended Newtonian dynamics with the Lagrangian $L=L(x,\dot{x%
},\ddot{x},\dddot{x},...,\dot{x}^{(n)})$ and Newtonian dynamics in the
inertial reference frames with the Lagrangian $L=L(x,\dot{x})$:

\begin{equation}
\int (L(x,\dot{x},\ddot{x},\dddot{x},...,\dot{x}^{(n)})-L(x,\dot{x}))dt=S(x,%
\dot{x},\ddot{x},\dddot{x},...,\dot{x}^{(n)})-S(x,\dot{x})=h  \label{12}
\end{equation}

Let us apply the least action principle [5]:

\begin{equation}
\delta S=\delta \int L(\dot{r^{\prime }},r^{\prime })dt=\int
\sum_{n=0}^{N}(-1)^{n}\frac{d^{n}}{dt^{n}}\frac{\partial L}{\partial \dot{r}%
^{(n)}}\delta \dot{r}^{(n)}dt=0.  \label{13}
\end{equation}

Then the generalized Euler-Lagrange equation for real reference frames will
take on the form
\begin{equation}
\sum_{n=0}^{N}(-1)^{N}\frac{d^{N}}{dt^{N}}\frac{\partial L}{\partial \dot{r}%
^{(N)}}=0.  \label{14}
\end{equation}

Or,
\begin{equation}
\frac{\partial L}{\partial r}-\frac{d}{dt}\frac{\partial L}{\partial \dot{r}}%
+\frac{d^{2}}{dt^{2}}\frac{\partial L}{\partial \ddot{r}}-...+(-1)^{N}\frac{%
d^{N}}{dt^{N}}\frac{\partial L}{\partial \dot{r}^{(N)}}=0.  \label{15}
\end{equation}

Let us generalize the above for the case of curvilinear coordinates. To do
so, one has to take into account the fact that in case of parallel
translation of a vector along non-straight trajectory of the body, not only
its value could be altered in the curved space, but as well its direction.
Therefore, applying covariant derivative in parameter $\tau $ of the vector $%
A^{a},a=1,2,3$

\begin{center}
$\nabla _{i}A^{k}=\frac{\partial A^{k}}{\partial x^{i}}+\Gamma
_{ij}^{k}A^{j} $
\end{center}

Let us introduce the operator

\begin{center}
$\dot{D}^{(1)}=\frac{dx^{i}}{d\tau }\nabla _{i}=\frac{D}{D\tau }$

$\dot{D}^{(2)}A^{i}=\frac{\partial \dot{D}^{(1)}A^{i}}{\partial \tau }%
+\Gamma _{jk}^{i}\dot{D}^{(1)}A^{j}\frac{dx^{k}}{d\tau }$

$...$

$\dot{D}^{(N)}A^{i}=\frac{\partial \dot{D}^{(N-1)}A^{i}}{\partial \tau }%
+\Gamma _{jk}^{i}\dot{D}^{(N-1)}A^{j}\frac{dx^{k}}{d\tau }$
\end{center}

Then the generalized Euler-Lagrange equations will take on the form

\begin{center}
$\frac{\partial L}{\partial r}-\frac{d}{dt}\frac{\partial L}{\partial \dot{D}%
^{(1)}r}+\frac{d^{2}}{dt^{2}}\frac{\partial L}{\partial \dot{D}^{(2)}r}%
-...+(-1)^{N}\frac{d^{N}}{dt^{N}}\frac{\partial L}{\partial \dot{D}^{(N)}r}%
=0 $
\end{center}

When we develop the proposed model, we have to employ functions implementing
stochastic variables. In particular, this is necessary to consider the
stochastic phase of oscillation that may be caused by stochastic fields. As
the nature of these fields is unknown, let us illustrate this case on the
example of a physical model with a stochastic gravitational/inertial
background (for non-inertial reference frames) and a distribution function
assuming uniform distribution of these fields in time and space. This means
that we suggest illustrating fluctuations of gravitational/inertial fields
and waves mathematically expressed by a stochastic curved space.

Then, considering quantum microobjects in the curved space, we must take
into account the fact that the scalar product of two 4-vectors $A^{i}$ and $%
B^{k}$ is $g_{ik}A^{i}B^{k}$ , where for weak gravitational fields one may
use the value $h_{ik}$, which is the solution of Einstein's equations for
the case of weak gravitational field in harmonic coordinates.

The correlation factor $M$ of the projection of stochastic vector variables $%
\lambda ^{i}$ onto directions $A^{k}$ and $B^{n}$ set by the polarizers (all
these vectors being unity ones) is [6]

\begin{center}
$\left\vert M\right\vert =\left\vert <AB>\right\vert =\left\vert <\lambda
^{i}A^{k}g_{lk}\lambda ^{m}B^{n}g_{mn}>\right\vert =\left\vert \frac{1}{2\pi
}\int \cos \phi \cos (\phi +\theta )d\phi \right\vert =$

$=\left\vert \cos \theta \right\vert $
\end{center}

due to the equations following from differential geometry,

\begin{center}
$\cos \phi =\frac{g_{ik}\lambda ^{i}A^{k}}{\sqrt{\lambda ^{i}\lambda _{i}}%
\sqrt{A^{k}A_{k}}},$

$\cos (\phi +\theta )=\frac{g_{mn}\lambda ^{m}A^{n}}{\sqrt{\lambda
^{m}\lambda _{m}}\sqrt{B^{n}B_{n}}}$.
\end{center}

Here $\phi $ is the angle between $\lambda ^{i}$ and $A^{k}$, $(\phi +\theta
)$ is between $\lambda ^{m}$and $B^{n}$.

This means coincidence of the Bell's observable [7] with the experimental
results in real (quantum) reference frames. All vectors here being unity
ones with metrics averaging in the weak field approximation yielding unity;
is the angle between polarizers, vector $A^{n}$ is equal to vector $B^{n}$
rotated by the angle $\theta $, and indices taking the values $i=0,1,2,3$.
Finally, we get
\begin{equation}
\left\vert M_{AB}\right\vert =\left\vert \cos \theta \right\vert .
\label{16}
\end{equation}

The maximum value of the Bell's observable $S$ is

\begin{center}
$\left\vert <S>\right\vert =\frac{1}{2}\left\vert <M_{AB}>+<M_{A^{\prime
}B}>+<M_{AB^{\prime }}>-<M_{A^{\prime }B^{\prime }}>\right\vert =$

=$\frac{1}{2}\left\vert \cos (-\frac{\pi }{4})+\cos (\frac{\pi }{4})+\cos (%
\frac{\pi }{4})-\cos (\frac{3\pi }{4})\right\vert =\sqrt{2}$,
\end{center}

where $\theta =\frac{\pi }{4}$ , being the doubled angle between direction
of the polarizers $A$ and $B$.

To sum it up, all equations of classical mechanics in the proposed model
possess additional terms in the form of higher derivatives. At that, these
additional terms are zero not always, but only in special cases, i.e. in the
inertial reference frames.

Additional terms in the form of higher derivatives may play the role of
hidden variables complementing both quantum and classic mechanics.
Additional terms have non-local character, which enables their employment
for description of non-local effects of quantum mechanics.

\end{document}